\documentclass[preprint,12pt]{elsarticle}

\usepackage[T1]{fontenc}
\usepackage{color}
\usepackage{verbatim}
\usepackage{float}
\usepackage{mathrsfs}
\usepackage{amsmath}
\usepackage{amssymb}
\usepackage{graphicx}
\usepackage{wasysym}

\newcounter{bla}

\journal{Machine Learning: Science and Technology}

\begin{document}

\begin{frontmatter}



\title{JefiAtten: An Attention-Based Neural Network Model for Solving Maxwell's Equations with Charge and Current Sources}


\author[a]{Ming-Yan Sun}
\author[a]{Peng Xu}
\author[b]{Jun-Jie Zhang\corref{author1}}
\author[b]{Tai-Jiao Du}
\author[c]{Jian-Guo Wang\corref{author2}}

\cortext[author1] {\textit{E-mail address:} zjacob@mail.ustc.edu.cn.}
\cortext[author2] {\textit{E-mail address:} wanguiuc@mail.xjtu.edu.cn}

\address[a]{Xi'an Research Institute of High Tech, Xi'an 710021, China}
\address[b]{Division of Computational Physics and Intelligent Modeling, Northwest Institute of Nuclear Technology, Xi'an 710024, China}
\address[c]{School of Information and Communications Engineering, Xi'an Jiaotong University, Xi'an 710049, China}

\begin{abstract}
We present JefiAtten, a novel neural network model employing the attention mechanism to solve Maxwell's equations efficiently. JefiAtten uses self-attention and cross-attention modules to understand the interplay between charge density, current density, and electromagnetic fields. Our results indicate that JefiAtten can generalize well to a range of scenarios, maintaining accuracy across various spatial distribution and handling amplitude variations. The model showcases an improvement in computation speed after training, compared to traditional integral methods. The adaptability of the model suggests potential for broader applications in computational physics, with further refinements to enhance its predictive capabilities and computational efficiency. Our work is a testament to the efficacy of integrating attention mechanisms with numerical simulations, marking a step forward in the quest for data-driven solutions to physical phenomena.

\end{abstract}


\end{frontmatter}

\section{Introduction}

Plasma, being the fourth state of matter apart from solid, liquid, and gas, has garnered extensive attention from researchers due to its complex dynamics beyond various technological applications \cite{nicholson1983introduction}. Given the intrinsic properties of plasma, the Maxwell's equations occupy a pivotal role in plasma simulation. These electromagnetic equations, when integrated with additional physical models—such as fluid\cite{plasma_fluid,plasma_fluid1}, collisional\cite{plasma_collisional,plasma_collisional1} and kinetic models—to create sophisticated simulation frameworks that are particle-based (such as particle-in-cell (PIC) methods\cite{Particle_in_cell,Particle_in_cell1,PIC-chen,PIC-chen1}) or field-based (such as magneto-hydrodynamic (MHD) simulations\cite{Magnetohydrodynamics,Magnetohydrodynamics1}.) In the simulation of complex plasma systems, the speed and accuracy of solving Maxwell's equations reflect the capability of the complex plasma system to a certain extent. The classical Maxwell solvers, such as the Finite Element Method (FEM) \cite{johnson2012} and Finite Difference Time Domain (FDTD)\cite{KaneYee,Warren,Oskooi}, are esteemed for their high computational accuracy and extensive applicability. 

With the advent of the deep learning, it is natural to ask if we can use these tools to either increase the simulation performance or tackle challenging tasks? In an effort to surmount this question, deep learning has been applied by researchers as a tool for solving all sorts of scientific problems \cite{wang2023scientific}. In terms of equation related scenarios, the networks utilized are principally classified into three categories: data-driven (e.g., graph neural networks (GNNs)\cite{Kuhn_Repn_Rockstuhl_2023}, Convolutional neural network (CNN)\cite{,data_drivenPuzyrev}, 3D Earth-specific transformer (3DEST)\cite{Bi2023AccurateMG}, etc.), model-driven (e.g., Physics-Informed Neural Networks (PINNs)\cite{Raissi2019}, Fourier Neural Operators (FNOs)\cite{Li_Kovachk}, Deep Operator Networks (DeepONet)\cite{Sifan2021}, etc.), and the hybrid-approach (e.g., hybrid data- and physics-augmented convolutional neural network (WaveY-Net)\cite{Chen_Lupoiu_Mao_Huang_Jiang_Lalanne_Fan_2021},  hybrid modulation network (HMN)\cite{Zhao2022AFF}, etc.).

Inspired by the success of large language models in areas such as natural language processing and computer vision\cite{Tetko_Karpov_Van,ThomasWolf2020,9716741,Liu_2021_ICCV}, numerous researchers have extended their applications to the field of numerical simulation. A growing body of work has demonstrated the rapid solution of partial differential equations by employing attention mechanisms to decode inter-data relationships, thereby achieving precise outcomes\cite{Li2022TransformerFP,pmlr-v202-hao23c,liu2023mitigating}. These researches offer valuable insights for our own endeavors in solving Maxwell's equations.

In this paper, we propose "JefiAtten", a network model designed to solve Maxwell's equations. The core of the model is an attention mechanism that is adept at learning the interconnections among data elements. We show that JefiAtten can effectively handle the Maxwell's equations in 3D space-time with various configurations. Meanwhile, we demonstrate that JefiAtten, when trained, acquires accelerated computational speed over the classical integral methods while remaining the generalization capabilities. This generalization capability is manifested in two levels: 1. use the trained JefiAtten to predict a different scenario; 2. re-train JefiAtten for the first several time steps and use it to predict the successive time steps. Remarkably, we find that when the neural network encounters completely new scenarios that exceed the numerical scope of the training dataset, we only need to re-train the network with a few similar scenarios so that JefiAtten can adapt to these new configurations.

The paper is organized in the following structure. In Sec. \ref{sec:Background}, we present a short introduction of the Jefimenko's equations and the self-attention mechanism. In Sec. \ref{sec:STRUCTURE_OF_THE_MODEL}, we first introduce the data and platform used during the model training process. Then, we delve into the model details. In Sec. \ref{sec:Experiment}, we address the model's computational speed, predictive capacity, and generalization ability, emphasizing its advantages in comparison to traditional algorithms. Through tests with various training datasets, we showcase the merits of JefiAtten in solving Maxwell's equations. The conclusion and outlook of JefiAtten is in Sec. \ref{Conclusion}.

\section{Background\label{sec:Background}}
In this section, we commence by presenting the high-dimensional Jefimenko's equations, which constitutes the primary problem that we aim to solve. Then, we explore the concept of the attention mechanism, a component that plays a pivotal role within our model's architecture.

\subsection{Jefimenko's equations}
 
The Jefimenko's equations are named after the physicist Oleg D. Jefimenko. Incorporating the principles of finite light speed and the relativistic effects on the delayed propagation of fields, the Jefimenko's equations yield the distributions of electric and magnetic fields through the specification of the spatial distribution of charge and current densities. This formulation provides an essential theoretical tool for the analysis of electromagnetic phenomena.

\begin{eqnarray}
\mathbf{E}(\mathbf{r},t) & = & \frac{1}{4\pi\epsilon_{0}}\int\left[\frac{\mathbf{r}-\mathbf{r}^{\prime}}{|\mathbf{r}-\mathbf{r}^{\prime}|^{3}}\rho(\mathbf{r}^{\prime},t_{r})\right.\nonumber \\
 &  & +\frac{\mathbf{r}-\mathbf{r}^{\prime}}{|\mathbf{r}-\mathbf{r}^{\prime}|^{2}}\text{\ensuremath{\frac{1}{c}}}\frac{\partial\rho(\mathbf{r}^{\prime},t_{r})}{\partial t}\nonumber \\
 &  & \left.-\frac{1}{|\mathbf{r}-\mathbf{r}^{\prime}|}\text{\ensuremath{\frac{1}{c^{2}}}}\frac{\partial\mathbf{J}(\mathbf{r}^{\prime},t_{r})}{\partial t}\right]d^{3}\mathbf{r}^{\prime}\label{eq:E}\\
\mathbf{B}(\mathbf{r},t) & = & -\frac{1}{4\pi\epsilon_{0}c^{2}}\int\left[\frac{\mathbf{r}-\mathbf{r}^{\prime}}{|\mathbf{r}-\mathbf{r}^{\prime}|^{3}}\times\mathbf{J}(\mathbf{r}^{\prime},t_{r})\right.\nonumber \\
 &  & +\left.\frac{\mathbf{r}-\mathbf{r}^{\prime}}{|\mathbf{r}-\mathbf{r}^{\prime}|^{2}}\times\text{\ensuremath{\frac{1}{c}}}\frac{\partial\mathbf{J}(\mathbf{r}^{\prime},t_{r})}{\partial t}\right]d^{3}\mathbf{r}^{\prime}\label{eq:B}\\
t_{r} & = & t-\frac{|\mathbf{r}-\mathbf{r}^{\prime}|}{c},\label{eq:tr}
\end{eqnarray}
where $\mathbf{\rho}$ and $\mathbf{J}$ represent the charge density and current density respectively, $\mathbf{E}$ and $\mathbf{B}$ signify the electric field and magnetic field at position $\mathbf{r}$ and time $t$. $t_{r}$ denotes the time delay.
As the generalized solutions of Maxwell's equations, Jefimenko's equations\cite{Jefimenko_1995} obviate the need for the specification of boundary conditions within the computational process. Instead, these equations necessitate only the inputs of current density and charge density to resolve the electric and magnetic fields, which make it more convenient for us to train the network solely based on the physical quantities $\mathbf{\rho}$, $\mathbf{J}$, $\mathbf{E}$ and $\mathbf{B}$. The Jefimenko's equations can be used with the employment of transport equations, such as the Boltzmann equation, so that we can arrive at a plasma simulator\cite{ZHANG2022108328,RBG_Maxwell}. The interconnections of $\mathbf{\rho}$, $\mathbf{J}$, $\mathbf{E}$ and $\mathbf{B}$ forge a pronounced interdependence between them, thereby amplifying the intricacies associated with the solution process, suggesting the usage of the attention mechanism in deep neural networks.

\subsection{Attention mechanism\label{sec:Attention_mechanism}}

\begin{figure}[H]
\begin{centering}
\includegraphics[scale=1]{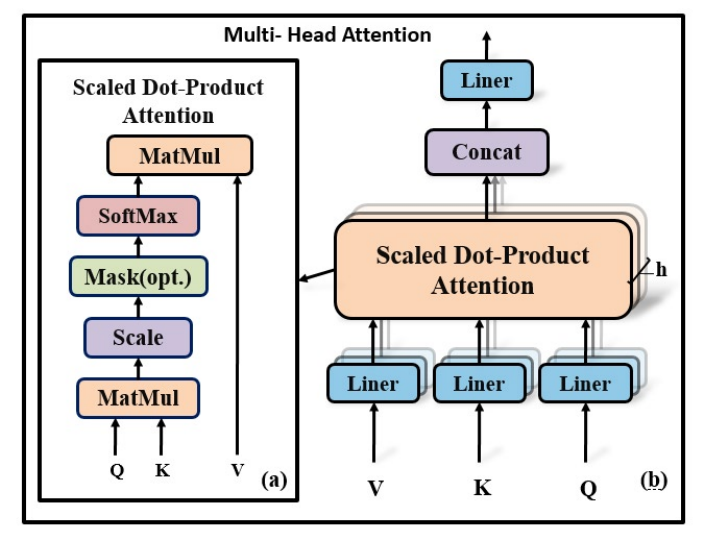}
\par\end{centering}
\caption{ (a) Scaled Dot-Product Attention; (b) The multi-head attention mechanism, composed of $h$ Scaled Dot-Product Attention modules. \label{fig:self-attention}}
\end{figure}

The top-down selective attention\cite{Tsotsos_Culhane_Kei1995} is characterized by its intentional nature and is task-specific in its deployment. Such a paradigm of selective attention can be efficaciously integrated into machine learning algorithms to mitigate the challenges posed by the deluge of information. A notable advancement was made in 2017 \cite{Vaswaniosukhin_2017}, which adeptly harnessed the attention mechanism to the realm of machine translation. The quintessence of the attention mechanism is encapsulated in its ability to compute the relational significance among data points. It segments the data into three fundamental constituents: query vectors $q_{j}$, key vectors $k_{j}$, and values $v_{j}$. The core computational process of this mechanism is depicted in Fig. \ref{fig:self-attention} a.

In attention mechanism, a linear mapping is firstly performed on the input vectors $q$, $k$, and $v$, so that these vectors are converted into the embedding space. Next, the similarity between $Q$ and $K$ ($Q = W^{Q}q$, $K = W^{K}k$, where $W^{Q}$ and $W^{K}$ are weight matrices) is obtained, followed by the Softmax normalization.
Finally, the obtained weights are used to perform a weighted summation on all the values in V ($Q = W^{V}v$).
\begin{equation}
\begin{array}{cc}
\text{Attention}(Q,K,V) = \text{softmax}(\frac{QK^{T}}{\sqrt{d_{k}}})V
\end{array}
\end{equation}

Due to the limitations of single attention operation in effectively handling complex tasks, a multi-head attention mechanism is introduced \cite{Vaswaniosukhin_2017}, as illustrated in Fig. \ref{fig:self-attention} b. 

This mechanism involves distributing the parameters into multiple subspaces, $ Q_{i} = QW^{Q}_{i},K_{i} = KW^{K}_{i},V_{i} = VW^{V}_{i}, i=1,...,n $ (where $n$ is the number of subspaces) where each subspace performs Scaled Dot-Product Attention independently. 
\begin{equation}
\begin{array}{cc}
\text{head}_{i}=\text{Attention}(Q_{i},K_{i},V_{i}),i=1,...,n 
\end{array}
\end{equation}
Owing to the independence of parameters within each distinct subspace, the multi-head attention mechanism surpasses a singular attention framework in its capacity to excavate information interrelations among data. Upon the completion of informational computations within the individual subspaces, the multi-head attention mechanism amalgamates the resultant outputs and subsequently conducts multiplication with a weight matrix.
\begin{equation}
\begin{array}{cc}
\text{MultiHead}(Q,K,V)=\text{Concact}(\text{head}_{1},...,\text{head}_{n})W^{O} 
\end{array}
\end{equation}

It is evident that the integration of attention mechanisms within neural networks substantially enhances their capacity to identify relationships among data. The depth of the network, established through the superposition of multi-layer attention mechanisms, is often instrumental in discerning the subtle and complex interrelations among data. Predicated upon this fundamental principle, we have engineered a model designed for the resolution of high-dimensional Maxwell's equations.

\section{Structure of the model\label{sec:STRUCTURE_OF_THE_MODEL}}

In this section, we first provide the requisite data and computational platform necessary for the training of the model. Subsequently, we present a detailed description of the model specifics.

\subsection{Data and Platform\label{sec:Data_and_Platform}}
The dataset employed in this training exercise was produced through the utilization of JefiGPU \cite{ZHANG2022108328}, our custom-developed, GPU-accelerated framework designed specifically for the computation of Jefimenko's equations. JefiGPU integrates discretized data with CUDA cores, harnessing the parallel processing of GPUs to expedite the calculation of Jefimenko's equations. Considering that the purpose of the training is to utilize Jefimenko’s equation in solving Maxwell's equations, it is essential for $\mathbf{\rho}$ and $\mathbf{J}$ to meet the $\nabla\cdot\mathbf{J}+\partial\rho/\partial t=0$ continuum relation.

To ensure the precision of the data, we have employed Mathematica \cite{Wolfram_1991} for the computation of $\mathbf{\rho}$ and $\mathbf{J}$. Due to the wide range of options for selecting $\mathbf{\rho}$ and $\mathbf{J}$, we aim to strike a balance between the complexity of the data and the computational time. To this end, we adopted the trigonometric functions as the basis functions and combined them in a flexible manner, resulting in a database comprising 500 evolution data sequences for model training and testing. The outcomes of these computations are accessible at the following repository: https://github.com/sunminmgyan/Data.git. 

We utilize the natural unit system for all mentioned physical quantities in the text,(we set $\hbar=c=k_{B}=\epsilon_{0}=1$) and the corresponding conversion relations can be found in Ref. \cite{UNITS}. We discretized the spatial domain into a $20\times20\times20$ grid and delineated the region containing $\rho$ and $\mathbf{J}$ as $[[-2.85, 2.85],[-2.85, 2.85],[-2.85, 2.85]] \text{ GeV}^{-3}$. The designated observation domain for the electric and magnetic fields was similarly confined to this 3D region. The temporal resolution was set with an interval, $dt$, of 0.005 $\text{GeV}^{-1}$. Electromagnetic field calculations were conducted using JefiGPU, with a running of 10,000 time steps, during which the values of $\rho$ and $\mathbf{J}$ were continuously updated by analytical equations. The details of the computational procedures are available for review within an example hosted at https://codeocean.com/capsule/0744578/tree/v1.

\begin{figure}[H]
\begin{centering}
\includegraphics[scale=0.46]{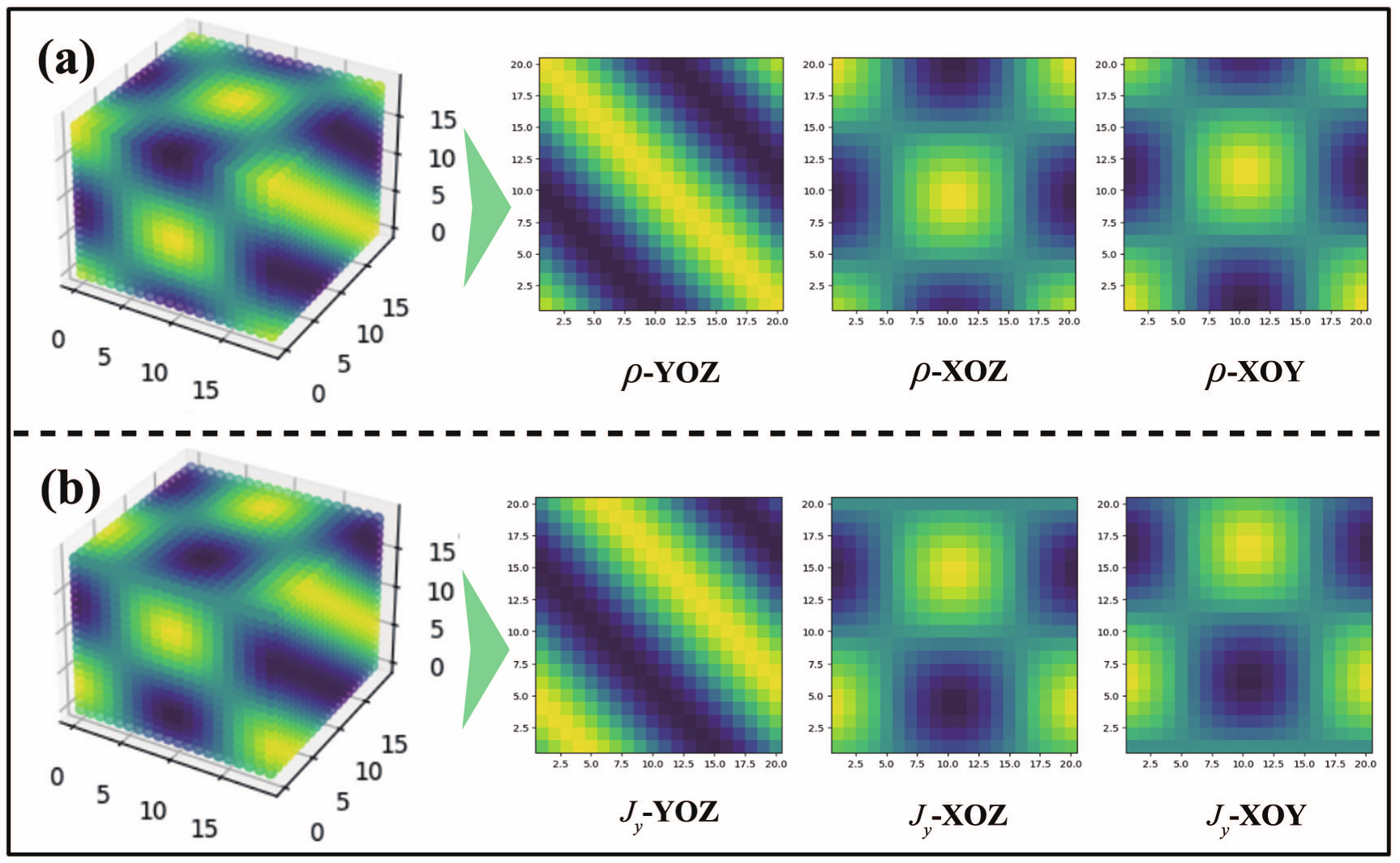}
\par\end{centering}
\caption{ Visualization of mesh of the sources $\rho$ and $\mathbf{J}$. (a) depicting the three-dimensional distribution of the charge density $\rho$, accompanied by the cross-sectional views of $\rho$ in the $x$, $y$, and $z$ directions; (b) illustrates the three-dimensional distribution of the current density $J_{y}$, with corresponding cross-sectional views of $J_{y}$ in the $x$, $y$, and $z$ directions.
\label{fig:data_show_rho_J}}
\end{figure}

\begin{figure}[H]
\begin{centering}
\includegraphics[scale=0.46]{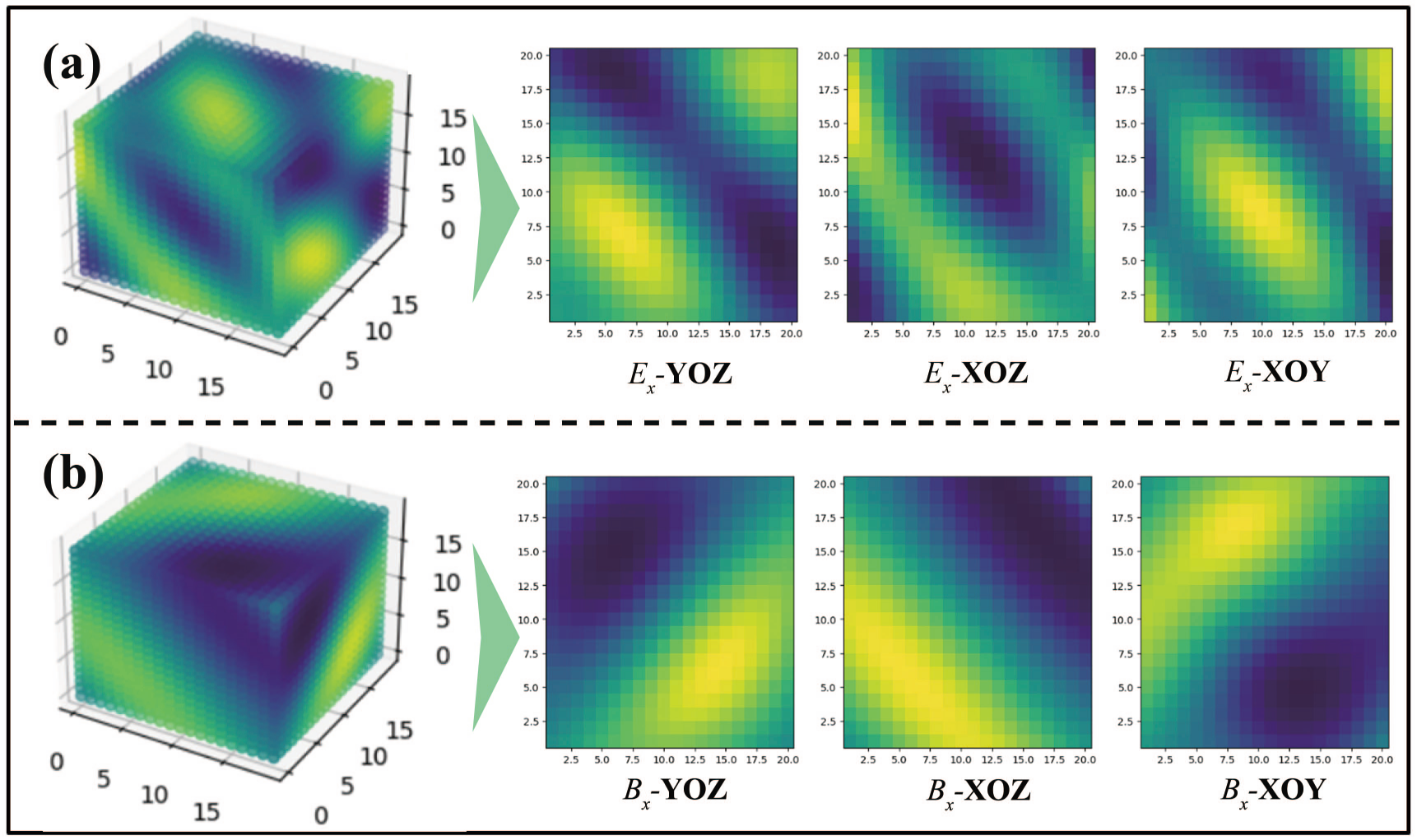}
\par\end{centering}
\caption{ Visualization of the electromagnetic fields $\mathbf{E}$ and $\mathbf{B}$. (a) depicting the three-dimensional distribution of the electric field $E_{x}$, accompanied by the cross-sectional views of $E_{x}$ in the $x$, $y$, and $z$ directions; (b) the three-dimensional distribution of the magnetic field $B_{x}$, with corresponding cross-sectional views of $B_{x}$ in the $x$, $y$, and $z$ directions.
\label{fig:data_show_E_B}}
\end{figure}

Beyond employing various combinations of trigonometric functions, we have also experimented with alterations to their amplitude and periodicity in order to evaluate the model's generalizability in the face of such changes. The computational process and the parameters remained consistent throughout; only the input values for $\mathbf{J}$ and $\rho$ were modified. The visualization of subsets from the dataset is shown in Fig. \ref{fig:data_show_rho_J} and Fig. \ref{fig:data_show_E_B}

In current work, we utilized four NVIDIA A100 cards for training the code and implemented parallel computing using the DataParallel package in PyTorch.

\subsection{ Overview of Model Architecture }

Here, we provide a succinct exposition of the model we have constructed. The Transformer architecture has gained considerable popularity in recent years, with numerous studies demonstrating its formidable capacity to process complex data and to extract information effectively. However, limitations of the Transformer persist in addressing multidimensional, highly-correlated, and structurally irregular datasets.

Due to the complexity of high-dimensional, time-dependent nature of the Maxwell equations, we have considered the potential impact of varying inputs on the network during the computational process. To address this, we have designed a fully connected neural network to preprocess the input data. This process is capable of not only processing the data across its multiple dimensions but also of preserving, to some extent, the inherent characteristics of the data. Diverging from the traditional Transformer architecture, we have removed the use of positional encoding to convey locational information within the data. Instead, we have instituted an attention module subsequent to the fully connected input network, which serves as a surrogate for the conventional positional encoding.

\begin{figure}[H]
\begin{centering}
\includegraphics[scale=0.11]{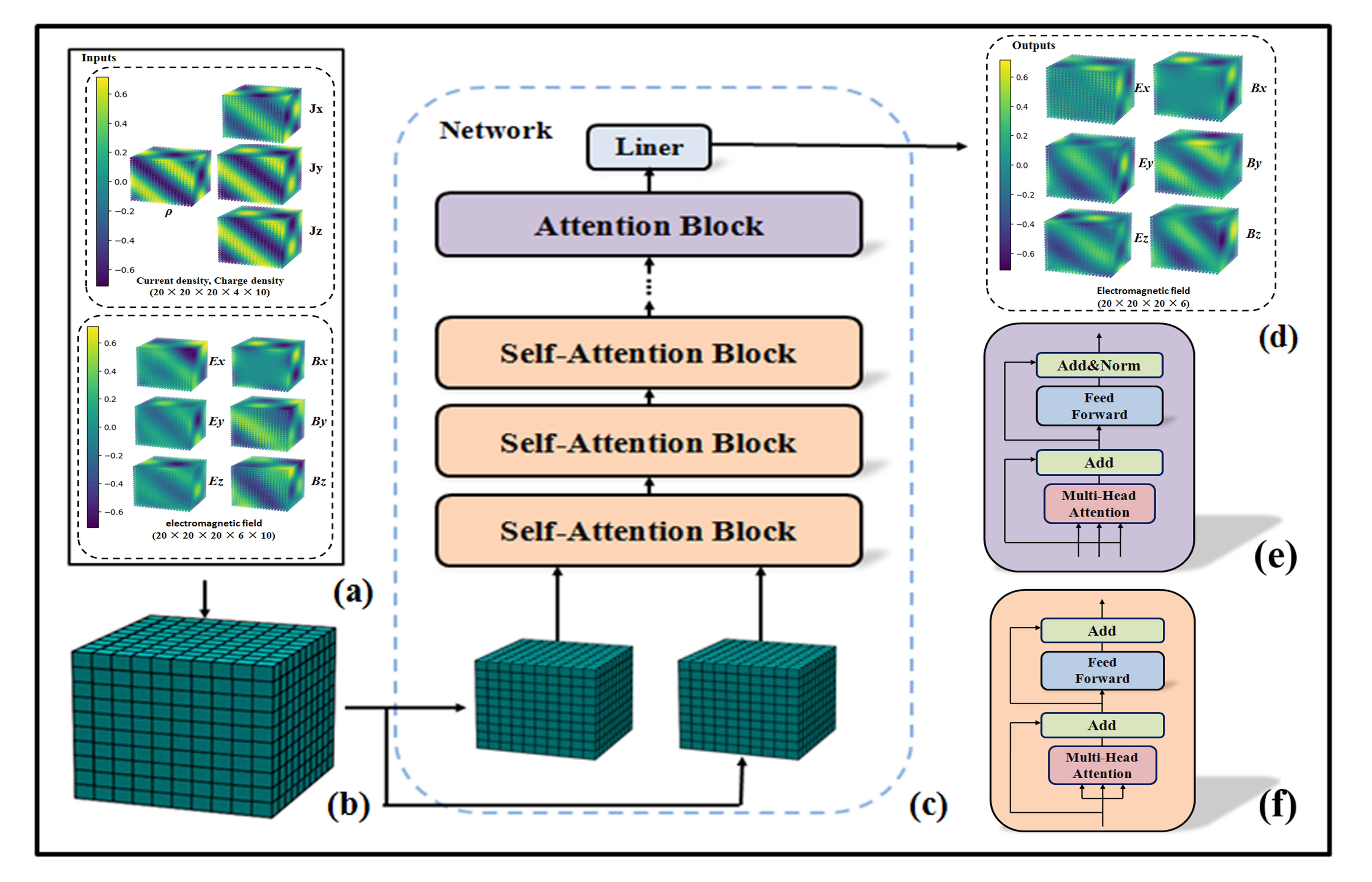}
\par\end{centering}
\caption{The JefiAtten's model structure. (a) Input current density and charge density at time $t_{1}$ and electromagnetic field at time $t_{1}$. (b) Grid partitioning of the input data. (c) The main structure of the network.(d) Output electric field and magnetic field at time $t_{2}=t_{1}+dt$. (e) Attention Block; (f) Self-attention Block. 
\label{fig:model-architcture}}
\end{figure}

Next, owing to the high degree of complexity inherent problems, a unitary network structure would be categorically insufficient for successful resolution. Inspired by the parallel computing approach, we use multiple attention modules in our network structure. The significance of the existence of these modules is to disassemble the original problem into different parts, and learn the correlation relationship in the same data by using independent parameters respectively, so as to achieve the purpose of fully mining the data relationship, and the structure of the whole model is shown in Fig. \ref{fig:model-architcture}.

\subsection{Attention module \label{sec:Attention_module}}

In our model, there are two types of attention modules, namely self-attention module and cross-attention module. The self-attention module is primarily used for learning the relationships within the data itself, such as the physical information, spatial information, and temporal information inherent in the data. On the other hand, the cross-attention module learns the relationships between the inputs $\rho$, $\mathbf{J}$, $\mathbf{E}$, and $\mathbf{B}$. The structure of each attention module resembles that of the Transformer model \cite{Vaswaniosukhin_2017},

\begin{equation}
\begin{array}{cc}
\begin{matrix}
f_{\text{middle}} =  \text{Atten}(f_{\text{input}}) + f_{\text{input}} \\
f_{\text{output}} = \text{LayerNorm}( FFN(f_{\text{middle}}) + f_{\text{middle}}),\label{eq:loss}
\end{matrix}
\end{array}
\end{equation}
where $Atten()$ represents the attention mechanism introduced in Sec. \ref{sec:Attention_mechanism}, and here we employ the multi-head attention mechanism. LayerNorm refers to layer normalization\cite{Ba_Kiros_Hinton_2016}, a technique that \cite{layerNorm} has demonstrated to confer stability to the input distribution during forward propagation within the network. Furthermore, it also contributes to the stabilization of the gradient during back-propagation. To prevent loss of information in the data, we only use LayerNorm at the end of cross-attention. $FFN()$ stands for the feed-forward neural network. If the model solely relies on the attention mechanism, the network would be limited to a linear expression. However, our purpose in constructing the network is to enable it to acquire more information from the data. Thus, the incorporation of non-linear features brought by FFN is crucial. Furthermore, the activation functions effectively transmit error signals during the back-propagation process, thus facilitating the construction of deeper networks. By augmenting the attention mechanism with FFN, we achieve a better fusion of linear and non-linear components, enabling more sophisticated data learning. Additionally, Ref. \cite{Geva_Schuster_Berant_Levy_2021} has also discovered the memory function of FFN in the field of NLP (Natural Language Processing), further validating its necessity in the network, which is the reason why we retain FFN in the attention module.

\subsection{ Training techniques }

To ensure the effectiveness of model learning, we incorporate a pre-training process before the formal training of the model. During pre-training, we use the input data at time $t$ and to predict the electromagnetic field at time $t+dt$. Additionally, we reduce the number of pre-training epochs. This approach has the advantage of using less time to obtain a set of initial weights for the entire network, thereby accelerating the training process of the entire task.

During the formal training process, to reduce the impact of accumulated errors and increase the network's performance, we combine $\rho$, $\mathbf{J}$, $\mathbf{E}$ and $\mathbf{B}$ at every ten time steps ($[t_{n-9},...,t_{n}]$) as the input to the model to predict the electromagnetic fields at $t_{n+1}$.  The predicted electromagnetic fields at time step $t_{n+1}$ is included as the input ($[t_{n-8},...,t_{n+1}]$) to predict the fields at time $t_{n+2}$, etc. This procedure increased the training dataset into ten times larger. While this training setup may increase the learning time of the model, it enhances the model's acceptance of errors, effectively improving its generalization ability. Furthermore, the updating process with standard data ensures that the model does not face the issue of being unable to fit the data.

We have selected the $\text{Smooth}L_{1}\text{Loss}$ function as our loss function, 
\begin{equation}
\begin{array}{cc}
\begin{matrix}
\text{Smooth}L_{1}\text{Loss}(f_{\text{output}} ,f_{\text{label}} ) \\= \frac{1}{n} {\textstyle \sum_{i=1}^{n}} \left\{\begin{matrix}
  0.5(f_{\text{output}} - f_{\text{label}} )^{2}, & \text{if }\begin{vmatrix}f_{\text{output}} - f_{\text{label}}\end{vmatrix}<1\\
  \begin{vmatrix}f_{\text{output}} - f_{\text{label}}\end{vmatrix} -0.5,& \text{otherwise}
\end{matrix}\right.
\end{matrix}
\end{array}
\end{equation}
which ensures that the gradient values do not become too large when there is a large discrepancy between the output values and the standard values. At the same time, it guarantees that the gradient values are sufficiently small when the output values are close to the standard values. Throughout the learning process, we utilize Adaptive Moment Estimation (Adam) to minimize the loss function and update the parameters of the entire network. Through the design of the network and optimization of the training process, our model is capable of adapting well to complex high-dimensional partial differential equations.
Due to the strong versatility of our model design, we can naturally extend the model to solve different types of partial differential equations by simply adjusting the parameters of the input and output.

\section{Experiment \label{sec:Experiment}}

In this section, we train and test the JefiAtten model on the established dataset. Our investigation focuses on assessing the model's computational precision, training duration, generalization capacity, and predictive efficacy.

\subsection{Model performance for scenarios with different $\rho$, $\mathbf{J}$ spatial distribution \label{sec:first_exp}}

We first apply the model to the dataset obtained using JefiGPU, as described in Sec. \ref{sec:Data_and_Platform}. During the training process, we use ten consecutive time steps ($[t_{n-9},...,t_{n}]$) as input, where for each time step the input consists of $\rho$ and $\mathbf{J}$, as well as $\mathbf{E}$ and $\mathbf{B}$ of the previous time step, e.g., [$\rho_{n}$, $\mathbf{J}_{n}$,  $\mathbf{E}_{n-1}$, $\mathbf{B}_{n-1}$]. The model is capable of predicting the next moment's $\mathbf{E}_{n}$ and $\mathbf{B}_{n}$ (or, if desired, predicting $\mathbf{E}$ and $\mathbf{B}$ for multiple future moments).

\begin{figure}[H]
\begin{centering}
\includegraphics[scale=0.55]{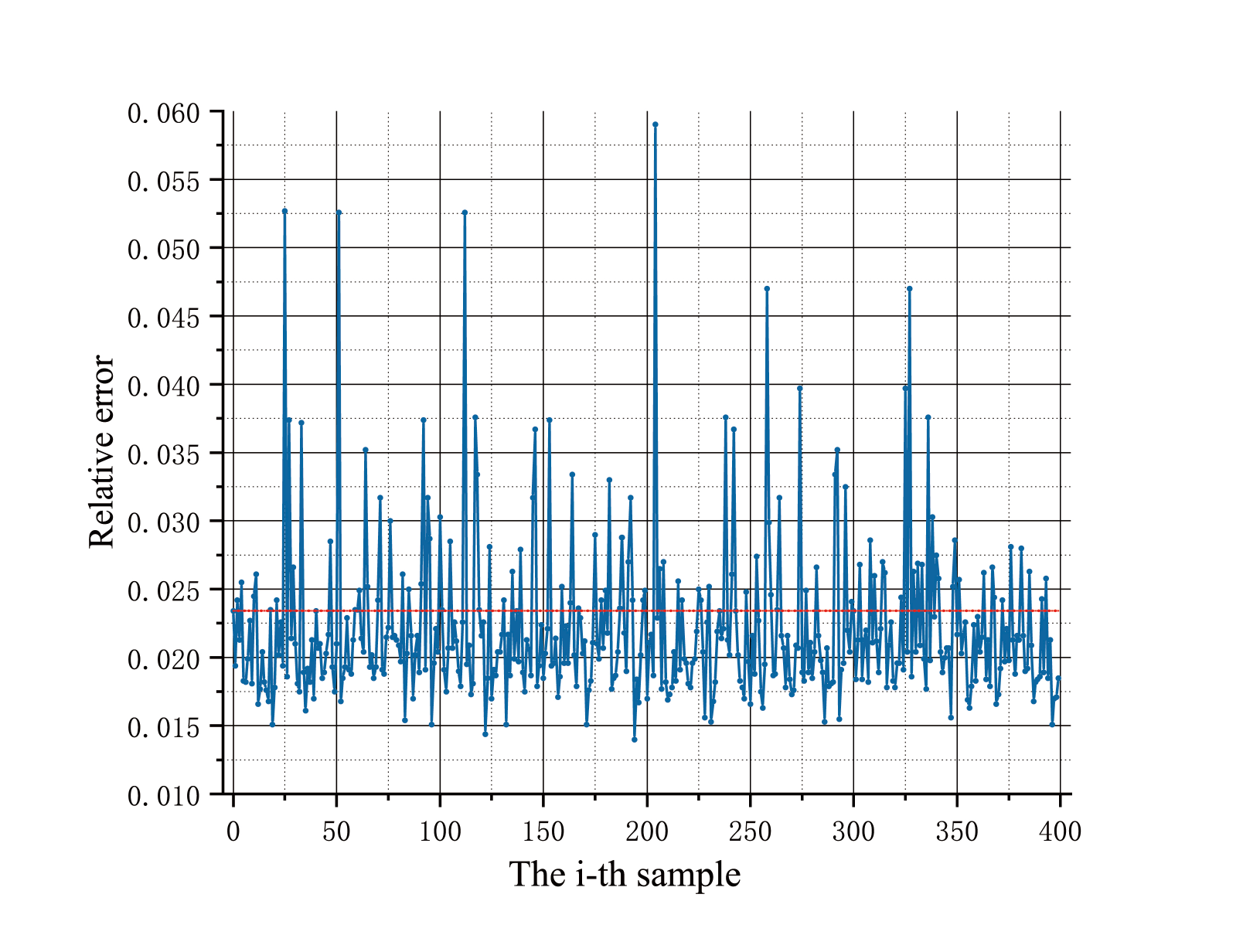}
\par\end{centering}
\caption{Relative error ($ \text{Mean}(\text{(Predicted-Ground-Truth)}^{2}) / \text{Mean}(\text{(Ground-Truth)}^{2}) $) between the predicted data and the ground-truth. The blue represents a group of 400 data points, while the red represents their mean. 
\label{fig:data-loss}}
\end{figure}

Here, we randomly choose 10 out of 500 time series for training, and utilize another randomly chosen 400 series for testing. In Sec. \ref{sec:Data_and_Platform}, we demonstrate that the data prepared possess varying spatial distribution, with substantial differences between them. Our purpose is to see if the network can correctly predict the electromagnetic fields of the testing samples with a few training samples, especially when these training and testing samples are very different in terms of spatial distribution. The result is depicted in Fig. \ref{fig:data-loss}.

In Fig. \ref{fig:data-loss}, we see that the maximum relative error of the 400 testing samples is about 0.06, and the average value stands at about 0.0234. With 10 training series, JefiAtten is able to grasp the intricate relationships between the input data $\rho$, $\mathbf{J}$, $\mathbf{E}$, $\mathbf{B}$. Meanwhile, JefiAtten boasts a faster computational speed compared to traditional algorithms once it has been trained. The training duration for JefiAtten stands at 3.34 hours (10 time series with each series having 1920 time steps; the total epoch number is taken to be 100). In the testing phase, for 2000 time steps on the same configuration, the traditional code JefiGPU requires 0.0542 hours on one NVIDIA A100 card, while JefiAtten completes the task in merely 0.0038 hours on the same device. 
The result indicates that JefiAtten is applicable in practical scenarios where, despite fixed boundary conditions and grid sizes, variations in current density and charge density occur. In traditional algorithms one would necessitate re-computation to adapt to these changes. In contrast, JefiAtten utilizes its  generalization capability to solve Jefimenko's equations, showcasing an advantage in terms of efficiency and utility.

\begin{figure}[H]
\begin{centering}
\includegraphics[scale=0.4]{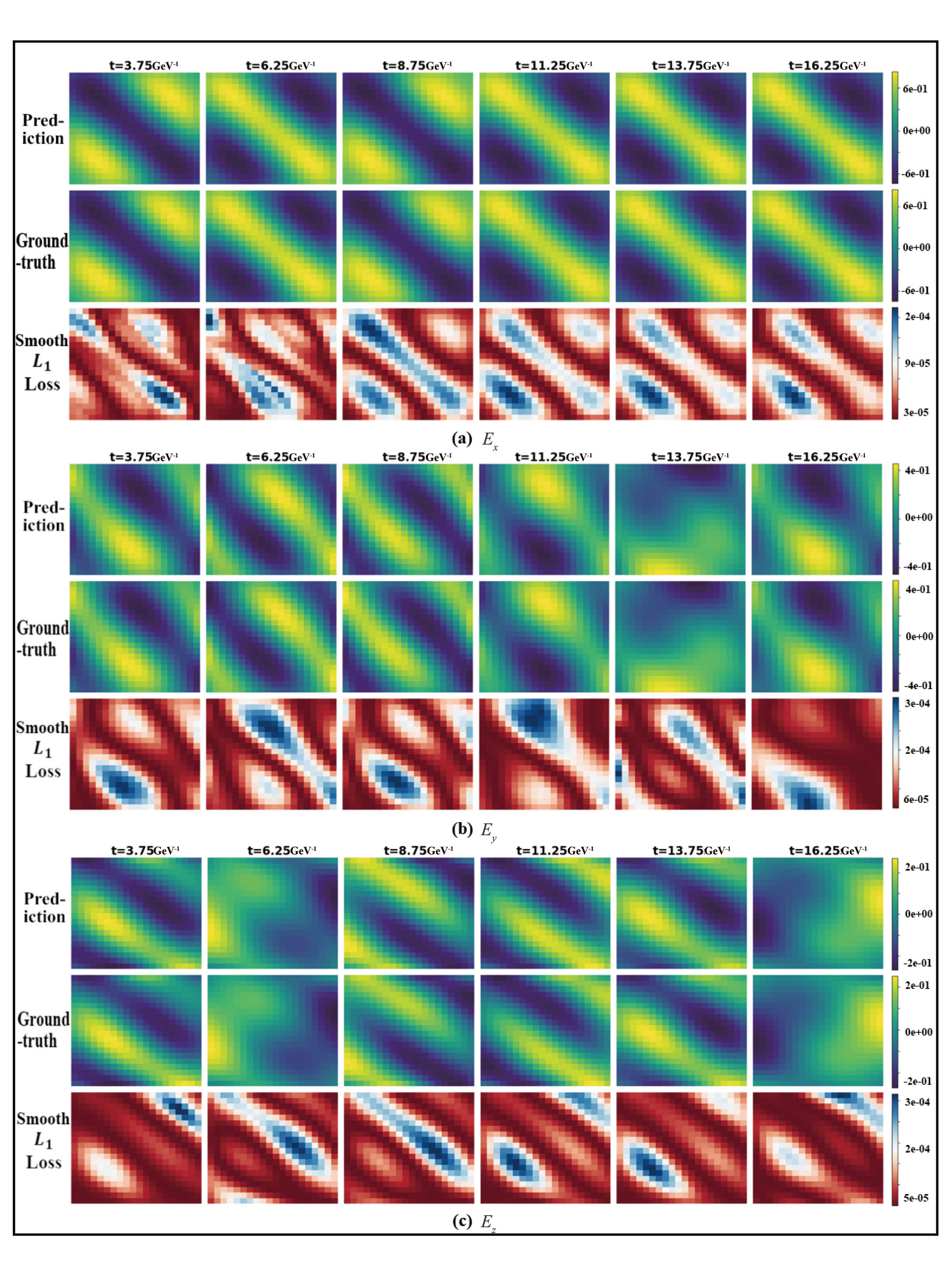}
\par\end{centering}
\caption{The predicted electric field and the absolute errors between the predictions and the ground-truth (at respective time steps of 3.75, 6.25, 8.75, 11.25, 13.75, 16.25 $\text{GeV}^{-1}$). (a) $E_{x}$ in the YOZ plane; (b) $E_{y}$ in the YOZ plane; (c) $E_{z}$ in the YOZ plane.
\label{fig:output1E}}
\end{figure}

\begin{figure}[H]
\begin{centering}
\includegraphics[scale=0.4]{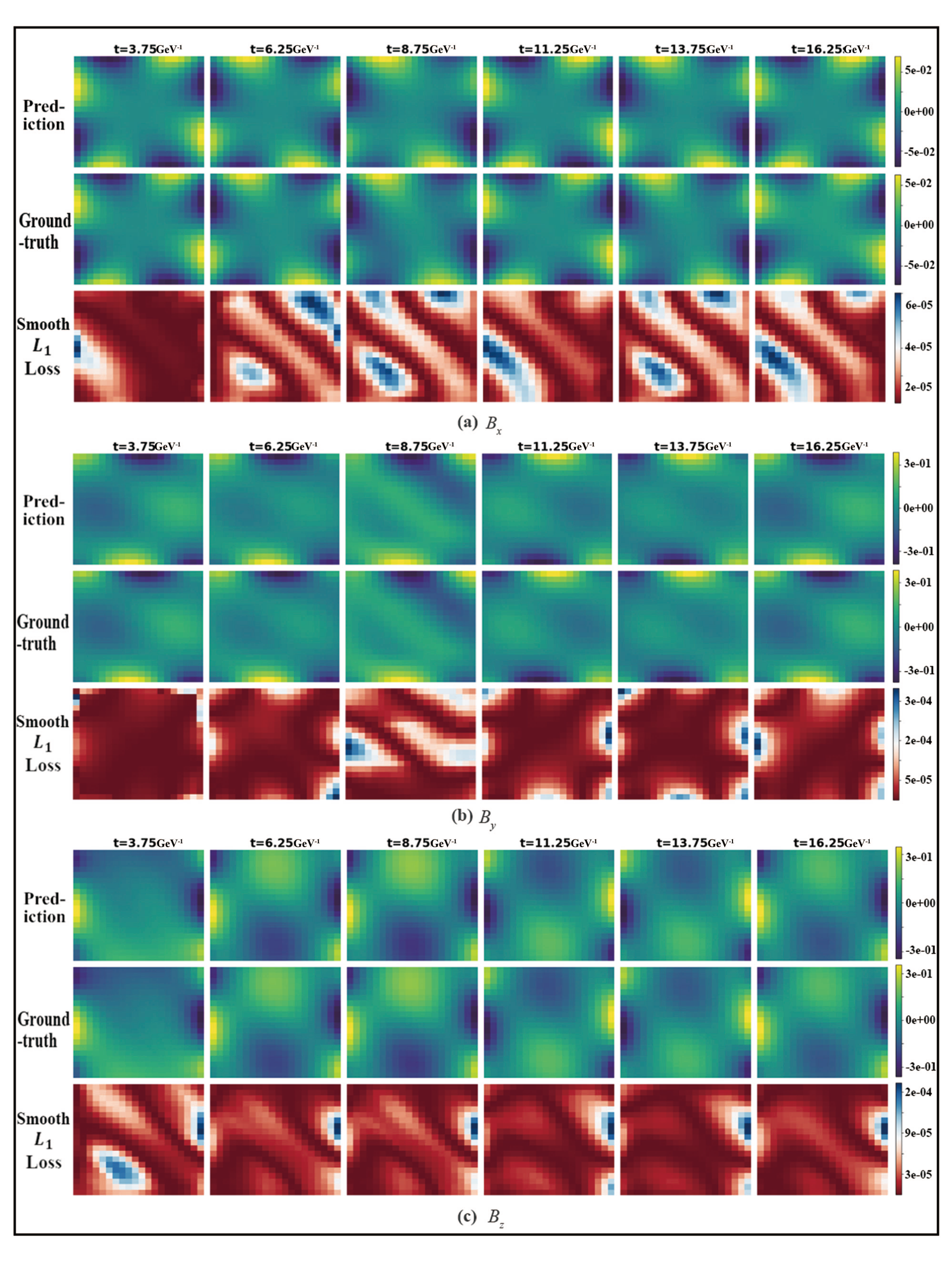}
\par\end{centering}
\caption{The predicted magnetic field and the absolute errors between the predictions and the ground-truth (at respective time steps of 3.75, 6.25, 8.75, 11.25, 13.75, 16.25 $\text{GeV}^{-1}$). (a) $B_{x}$ in the YOZ plane; (b) $B_{y}$ in the YOZ plane; (c) $B_{z}$ in the YOZ plane.
\label{fig:output1B}}
\end{figure}

To better illustrate the generalization ability, we visualize some of the predicted electromagnetic fields with corresponding errors in different directions. These results are showcased in Fig. \ref{fig:output1E} and Fig. \ref{fig:output1B}. From Fig. \ref{fig:output1E} and Fig. \ref{fig:output1B}, we see that the network predictions capture the main structures of the ground-truth. Given that the inputs $\rho$ and $\mathbf{J}$ are composed using trigonometric functions as basis functions, the predicted electromagnetic fields may exhibit certain periodicity, which is also evident in Fig. \ref{fig:output1E}.

Upon examining samples of the test data, we confirm that the model accurately fits the entire electromagnetic field without any partial fitting issues. For simplicity in presentation, we will focus on displaying only the $E_{x}$ component along the x-axis in upcoming results.

\subsection{Model performance for scenarios with different amplitudes}

In Sec. \ref{sec:first_exp}, we validated the model's computational accuracy for the Jefimenko's equations and its generalization to triangular functions with varying spatial distribution but consistent amplitude. Building upon the training discussed in Sec. \ref{sec:first_exp}, this section focuses on fine-tuning the network using a single time series with triangular amplitudes that are significantly larger. We then proceed to evaluate the model's ability to generalize to changes in amplitudes. The purpose of this section is to see network's capability when the trained network encounters scenarios that are very different from the training data, and how much new data is required for fine-tuning.

\begin{figure}[H]
\begin{centering}
\includegraphics[scale=0.65]{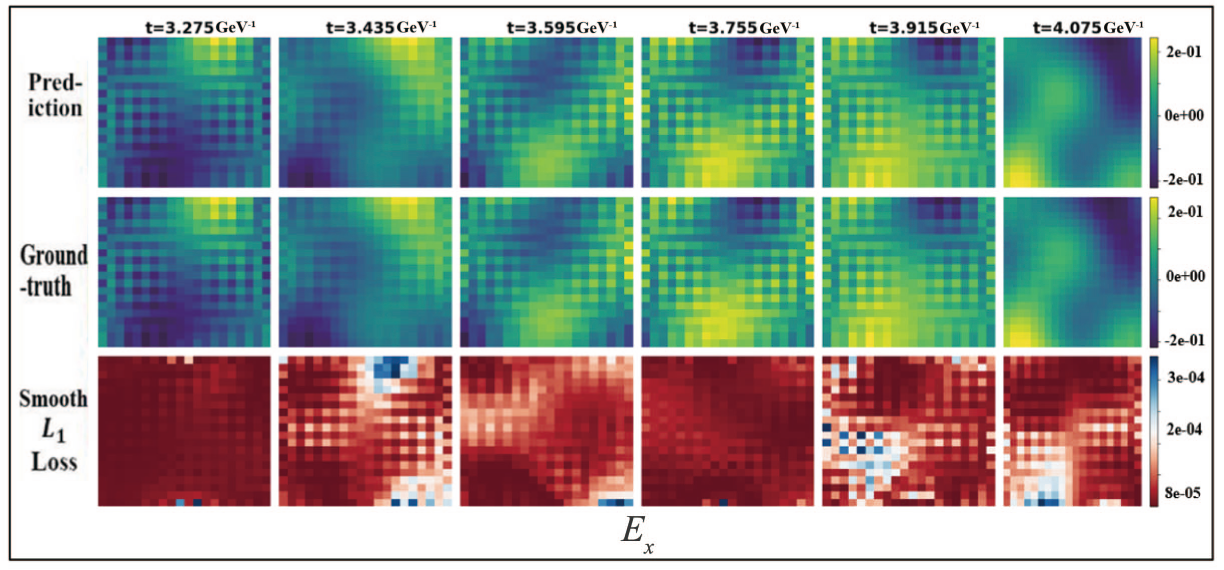}
\par\end{centering}
\caption{The predicted electric field and the absolute errors between the predictions and the ground-truth.(at respective time steps of 3.275, 3.435, 3.595, 3.755, 3.915, 4.075 $\text{GeV}^{-1}$)
\label{fig:same-amplitude-output-ex}}
\end{figure}

We begin by randomly selecting three time series from a batch of 500, amplifying their amplitudes to ten times their original levels. We designated one series for training and the other two for testing. Continued training of the model, leveraging the weights obtained in Sec. \ref{sec:first_exp}, serve to decrease the amount of time needed for fine-tuning. The results of the testing phase are detailed in Fig. \ref{fig:same-amplitude-output-ex}

The inputs for this set of results, $\rho$ and $\mathbf{J}$, are as follows:

\begin{equation}
\begin{array}{cc}
\rho (x,y,z,t) = \left\{ {\begin{array}{*{20}{c}}
{\rm{0}}&{{\rm{if\quad t < 0}}}\\
{10(\text{cos}(t)(\text{cos}(y)\text{cos}(x+10z)- \text{sin}(10y)}\\ { \text{sin}(x+z))+ \text{sin}(t)\text{sin}(y)\text{sin}(10x+z))}&{{\rm{if \quad t}} \ge {\rm{0}}}
\end{array}} \right.\\
\mathbf{J}(x,y,z,t) = \left\{ {\begin{array}{*{20}{c}}
{\rm{0}}&{{\rm{if\quad t < 0}}}\\
{({j_x},{j_y},{j_z})}&{{\rm{if\quad t}} \ge {\rm{0}}}\\
\end{array}} \right.
\label{eq:rhoandj1}
\end{array}
\end{equation}
where $j_x = \text{sin}(y)\text{cos}(10x+z)\text{cos}(t); \quad j_y = \text{cos}(10y)\text{sin}(x+z)\text{sin}(t); \quad j_z = \text{cos}(y)\text{sin}(x+10z)\text{sin}(t);$ Compared to the results presented in Sec. \ref{sec:first_exp}, this set of outputs exhibits a noticeable increase in complexity as well as larger amplitudes. However, as indicated by Fig. \ref{fig:same-amplitude-output-ex}, the model has captured the features compared to the ground-truth, with a maximum absolute error of $3\times10^{-4}$. These findings demonstrate the capability of JefiAtten to rapidly adapt to inputs of very different scenarios (varying amplitudes) based on its pre-training, and to accommodate more complex input conditions. This expands the potential applications of JefiAtten.

\begin{figure}[H]
\begin{centering}
\includegraphics[scale=0.5]{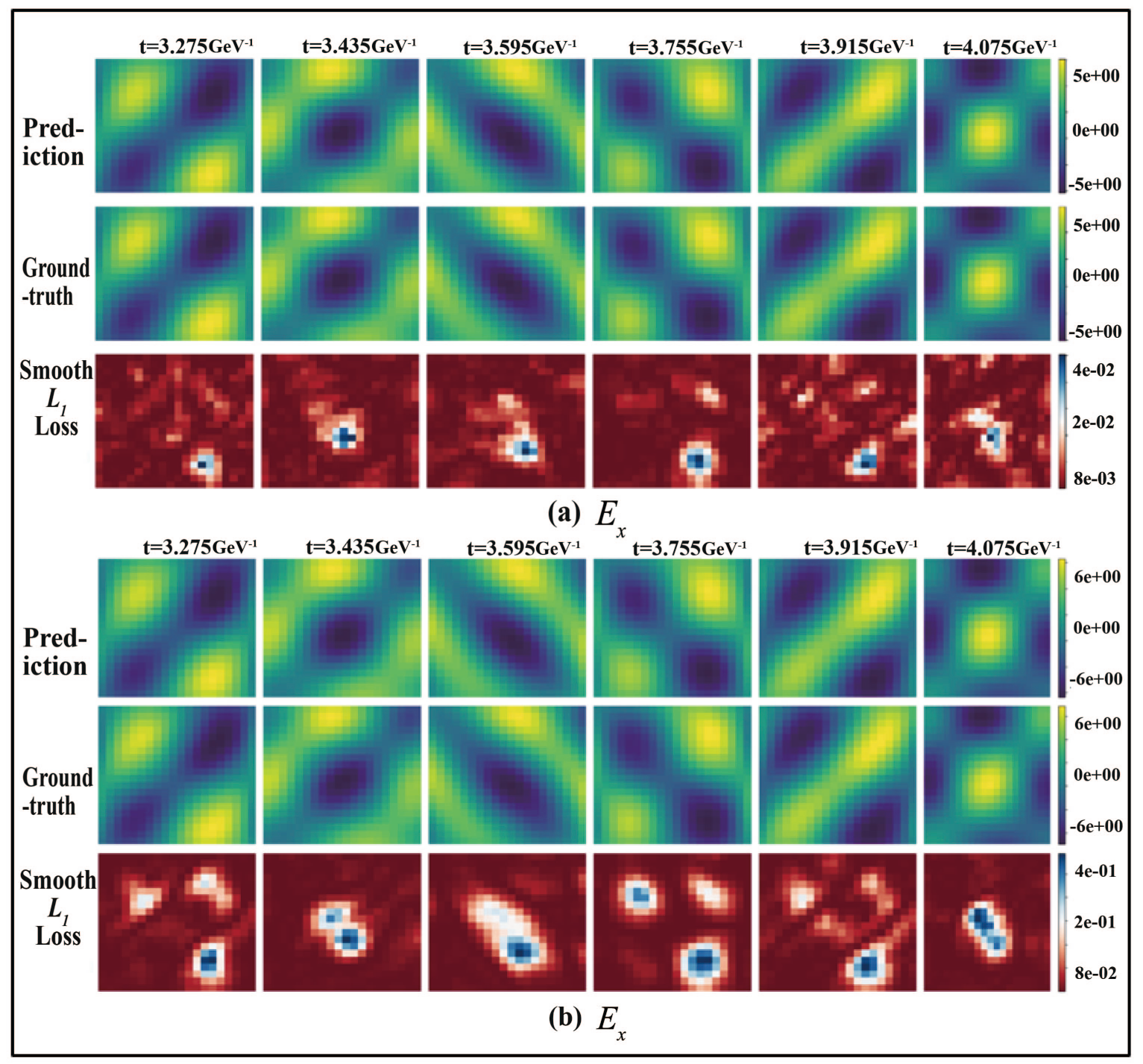}
\par\end{centering}
\caption{The predicted electric field and the absolute errors between the predictions and the ground-truth (at respective time points of 3.275, 3.435, 3.595, 3.755, 3.915, 4.075 $\text{GeV}^{-1}$). (a) $E_{x}$ in the YOZ plane. Set 1 and 3 used as fine-tuning, and set 2 used as testing (i.e., amplitude for training is 10 and 15, and for testing is 13); (b)$E_{x}$ in the YOZ plane. Set 1 and 5 used as fine-tuning, and set 4 used as testing (i.e., amplitude for training is 10 and 18, and for testing is 15).
\label{fig:differenet-range-ouput-Ex}}
\end{figure}

We continue to probe the limits of JefiAtten's generalization capabilities. As previously ascertained, JefiAtten has demonstrated generalization for amplitudes, and spatial distribution more or less present in the training set. Next, we test its performance with amplitudes that have never appeared during the training process. From a pool of 500 datasets, we select five sets with similar spatial distribution, and augment their amplitudes by factors of 10, 13, 15, 18, and 30, respectively (denoted as set 1-5). We use sets 1 and 3, as well as 1 and 5 for training, and sets 2 and 4 for testing, the obtained results are illustrated in Fig. \ref{fig:differenet-range-ouput-Ex}

The analysis of the figure suggests that the difference between the predictions of set 4 and the ground-truth values is more pronounced than in set 2, which has a test error of 0.0147, while set 4 has a test error of 0.0067. Despite this, both sets of predictions confirm JefiAtten's generalization ability. This implies a direct relationship between the accuracy of the model's outputs and the amplitude discrepancy between the training and test datasets. Meanwhile, we have the flexibility to balance computational speed with precision, tailored to the specific needs. For a quicker training cycle, one could choose to train the model within a predefined amplitude range; conversely, to achieve higher accuracy in computations, training with datasets that have amplitudes similar to those of the test set would be advantageous. These findings further demonstrate JefiAtten's ability to quickly adapt to inputs of varying amplitudes, utilizing its pre-trained foundation.

\subsection{Model performance for slowly varying scenarios}

\begin{figure}[H]
\begin{centering}
\includegraphics[scale=0.13]{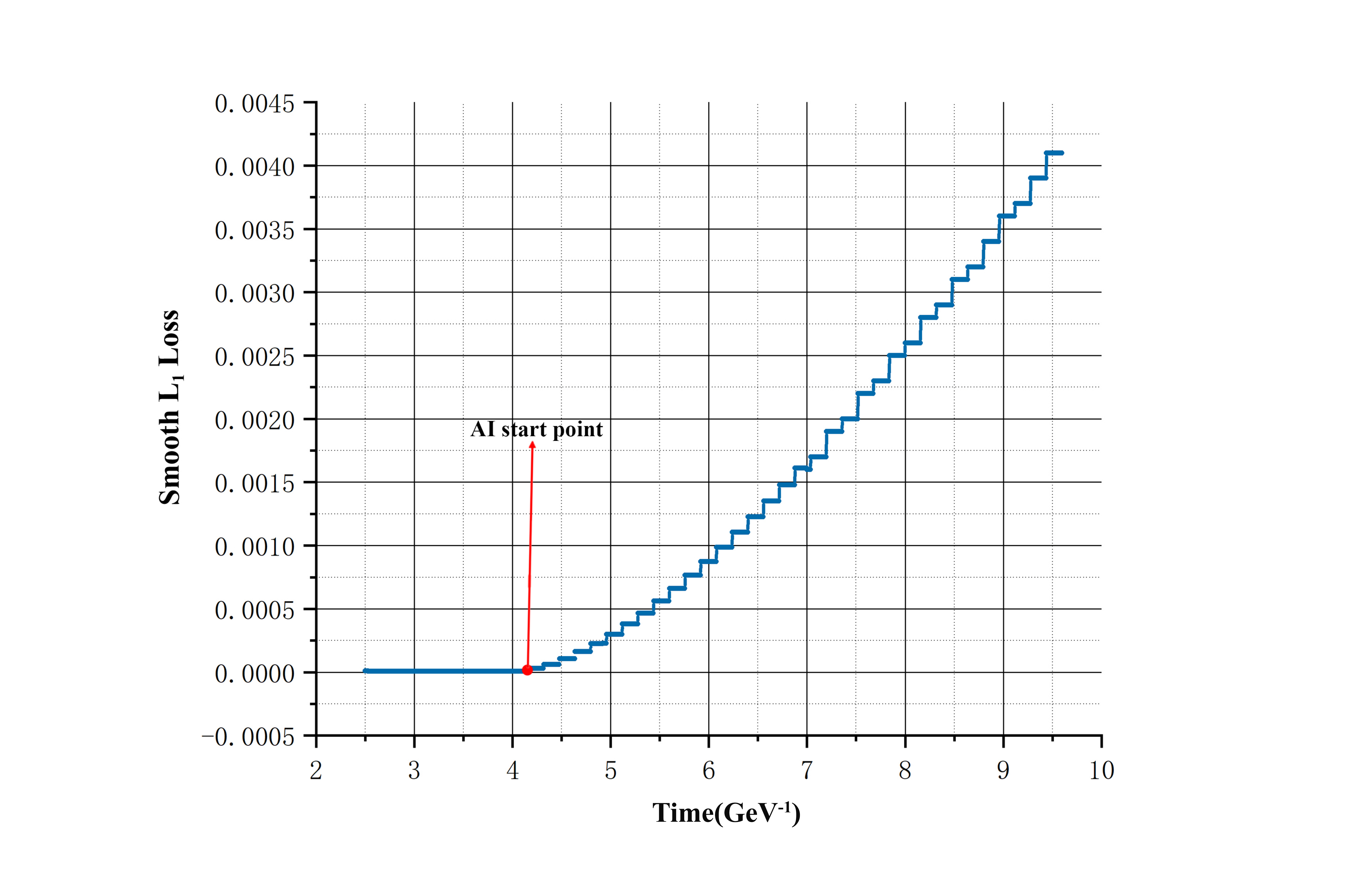}
\par\end{centering}
\caption{ The loss between the model's predicted value and the ground-truth value. The model predicts the system from the red dot, and the error is calculated using the SmoothL1Loss method, details of which can be found in Eq. \ref{eq:loss}
\label{fig:simple-function-anticipate1}}
\end{figure}

During the experimental process described above, we assessed the model's practicality, generalization ability, and computational speed from various perspectives. Ultimately, we aim to gain a deeper understanding of the model's predictive capabilities. Given the periodic nature of trigonometric functions, the model inevitably accumulates errors when making predictions on complex data, and this cumulative error can impact its predictive performance. However, when the input data has a simpler structure, the model tends to incur smaller errors during consecutive prediction (i.e., using the network to predict the subsequent electromagnetic fields based on its own predictions). Under these circumstances, the model should be capable of making predictions over a longer time-frame. Therefore, we set the input parameters $\rho$ and $\mathbf{J}$ as follows:

\begin{equation}
\begin{array}{cc}
\begin{matrix}
\rho(x,y,z,t)=\left\{\begin{matrix}
  0 & \text{if}\quad t<0 \\
  2(x+y+z)e^{-t} &  \text{if}\quad t\ge0
\end{matrix}\right. \\
\mathbf{J}(x,y,z,t) = \left\{\begin{matrix}
  0 & \text{if}\quad t<0 \\
  (j_{x},j_{y},j_{z}) &  \text{if}\quad t\ge0
\end{matrix}\right.
\end{matrix}
\end{array}
\end{equation}
where $j_x = (x^{2}+yz)e^{-t}, j_y = (y^{2}+xz)e^{-t}, j_z = (z^{2}+yx)e^{-t}.$ Based on the training conducted in  Sec. \ref{sec:first_exp}, we test this dataset, with the results displayed in Fig. \ref{fig:simple-function-anticipate1}. 

As depicted in Fig. \ref{fig:simple-function-anticipate1}, for simulated data with a time step of $dt=0.005 \text{ GeV}^{-1}$ , JefiAtten takes over the traditional equation solving process from 4s and predicts until 9.5 $\text{GeV}^{-1}$, covering a total of 1100 time steps. The maximum error generated by the predictions is 0.0043. This outcome further showcases the predictive ability of JefiAtten when dealing with simple functions.

\subsection{Ablation Experiments}
The JefiAtten model, as depicted in  Fig. \ref{fig:model-architcture}, did not originate in its current form. The iterative concept behind its design can be referred to in Appendix. In order to further investigate the rationality of the JefiAtten model's structure, we conducted ablation experiments targeting specific components within the model. The results of these experiments are presented in Tab. \ref{tab:Ablation}, shedding light on the significance of each component in the overall performance of the model.

\begin{table}
\caption{Experimental results for the necessity of different model blocks. \label{tab:Ablation}}
\centering{}
    \centering
    \begin{tabular}{|l|l|}
    \hline
        Network architecture & Deviation \\ \hline
        No inputattention & 0.2873 \\ \hline
        No FFN & 0.4050 \\ \hline
        Add Layernorm & 0.4165 \\ \hline
        No change & 0.0025 \\ \hline
    \end{tabular}
\end{table}

Among the ablation experiments conducted, "No inputattention" indicates that the model eliminates the attention module for capturing temporal and positional information from the input. "No FFN" denotes the removal of the Feed-Forward Network (FFN) in each module. "Add Layernorm" signifies the addition of Layer Normalization in each module. "No change" represents the JefiAtten model without any modifications. The results demonstrate that the attention module for temporal and positional information effectively captures the relevant information. Compared to position masking, this approach offers a broader application range, enabling rapid acquisition of the temporal and positional associations within complex input data. As described in Sec. \ref{sec:Attention_module}, the FFN component provides an effective means of expressing non-linearity, as confirmed by the results of the ablation experiments. On the other hand, the inclusion of Layer Normalization in each module can lead to a loss of information and an increase in the error of the final output results.

\section{Conclusion\label{Conclusion}}

In this paper, we introduced JefiAtten, a novel neural network model designed to solve Maxwell's equations through an attention-based mechanism. JefiAtten leverages the power of self-attention and cross-attention modules to capture intricate relationships within and between the data inputs, ensuring a comprehensive understanding of the physical, spatial, and temporal information embedded in the electromagnetic data. Our model's architecture adeptly incorporates fully connected networks for initial data preprocessing, followed by a series of attention blocks for detailed feature extraction. This design choice is validated by our ablation studies, which highlight the necessity of each component in achieving low prediction errors.

JefiAtten has shown improved performance in both pre-training and formal training phases, outpacing classical computational methods in terms of speed once adequately trained. The model's generalization capabilities are evidenced by its ability to adapt to new scenarios with varying spatial distribution, amplitudes, and slowly varying conditions with limited additional training. 

Our findings open up pathways for future research, where JefiAtten can be further enhanced to tackle even more complex simulations, potentially serving as an encouragement for further exploration into the fusion of machine learning with scientific computing.

\section{Acknowledgement}
The work is supported by the National Natural Science Foundation of China (NSFC) under the grant number 12105227 and the National Key Research and Development Program of China under Grant No. 2020YFA0709800.

\section*{Appendix}
\begin{figure}[H]
\begin{centering}
\includegraphics[scale=0.5]{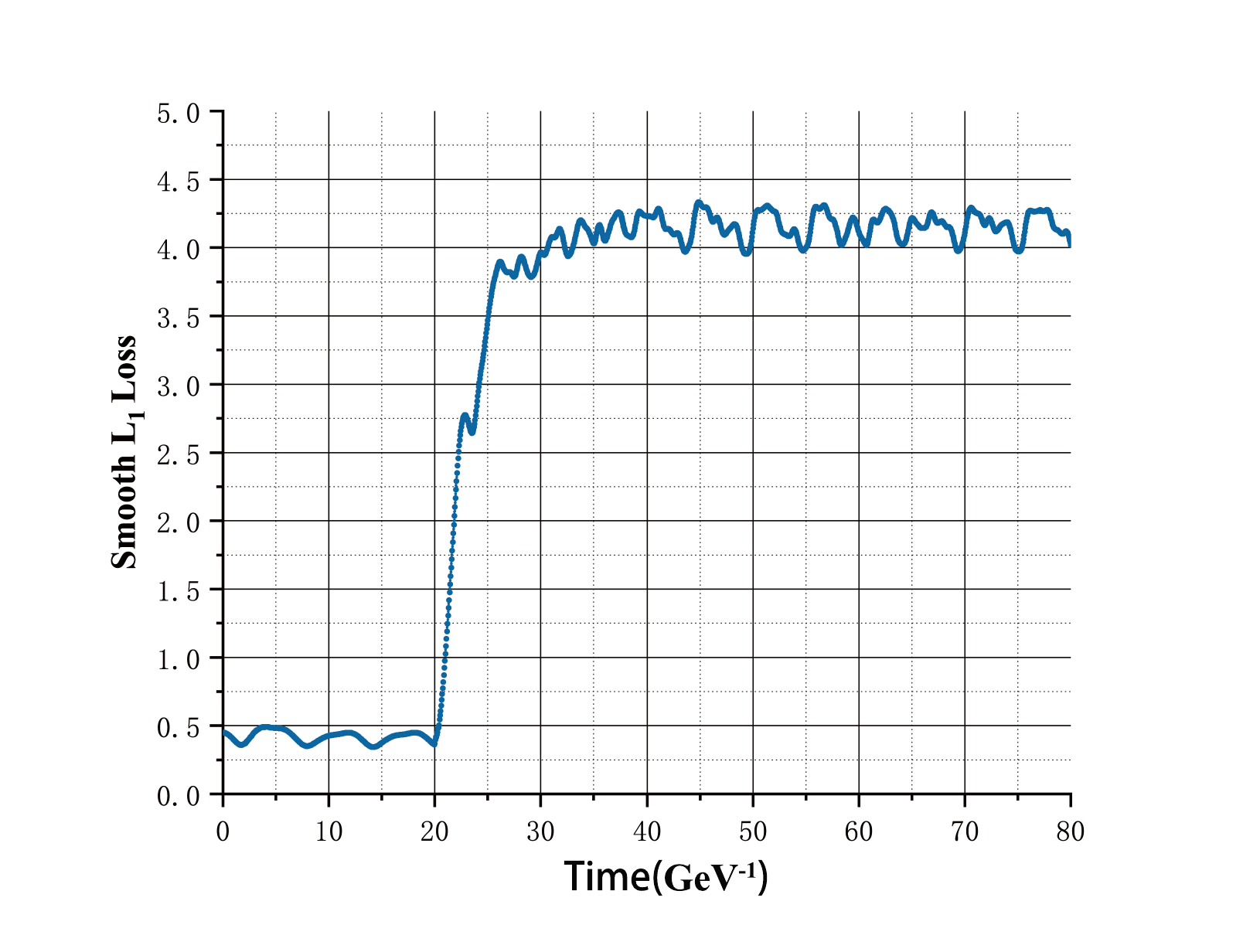}
\par\end{centering}
\caption{ The error between the model's predicted value and the Ground-Truth.
\label{fig:error_point}}
\end{figure}

It is intriguing to consider why we chose to utilize multiple stacked attention mechanisms, as depicted in  Fig. \ref{fig:model-architcture}, in our network architecture. From Eq. \ref{eq:E}, Eq. \ref{eq:B} (Jefimenko's equation), it can be observed that when computing the electromagnetic field, the high temporal requirements imposed by the presence of a time delay $t_{r}$ necessitate the model's ability to effectively capture the temporal relationships between different data points. This observation emerged during the initial stages of our model exploration. Initially, we trained the model using only the previous time-step's electromagnetic field, current density, and charge density as input data. However, after training, we noticed that while the model could accurately predict the electromagnetic field output for shorter time intervals, it performed poorly for longer time intervals, as illustrated in Fig. \ref{fig:error_point}.
Upon analyzing the reasons behind this phenomenon, we discovered that the model exhibited a heavy reliance on the input results from the previous time-step. During the prediction process, the model accumulated significant errors, leading to substantial deviations in the data. This is precisely why we combined multiple time series inputs and employed attention mechanisms to explore the inter-dependencies between the data.
Moreover, due to the high complexity of the input data ($E_{x}, E_{y}, E_{z}, B_{x}, B_{y}, B_{z}, rho, J_{x}, J_{y}, J_{z}, t $), integrating the positional information with the input data posed considerable challenges. Consequently, we incorporated the positional information as an input and utilized attention to identify the correlations between the positional information and the input data. This is also why our model adopted a structure involving multiple stacked attention mechanisms, as depicted in  Fig.  \ref{fig:model-architcture}. 

\section*{Declaration of Generative AI and AI-assisted technologies in the writing process}

During the preparation of this work the authors used GPT-4 in order to improve readability and language. After using this tool, the authors reviewed and edited the content as needed and take full responsibility for the content of the publication.

\bibliography{sample}

\bibliographystyle{elsarticle-num}

\end{document}